\documentstyle[epsfig,here,12pt]{article}

\newcommand {\eqref} [1] {(\ref {#1})}
\newcommand {\beq} {\begin{equation}} 
\newcommand {\eeq} {\end{equation}}
 \newcommand {\ber}{\begin{eqnarray*}}
 \newcommand {\eer} {\end{eqnarray*}}
\newcommand {\bea}{\begin{eqnarray}}
 \newcommand {\eea} {\end{eqnarray}} 

\newcommand{\Nfour} {${\cal N}=4\ $}
\newcommand{\Ntwo}{${\cal N}=2\ $}
\newcommand{\None}{${\cal N}=1\ $}
\newcommand {\state}
[1] {\mid \! \! {#1} \rangle} 
\newcommand{\drawsquare}[2]{\hbox{%
\rule{#2pt}{#1pt}\hskip-#2pt%  left vertical
\rule{#1pt}{#2pt}\hskip-#1pt%  lower horizontal
\rule[#1pt]{#1pt}{#2pt}}\rule[#1pt]{#2pt}{#2pt}\hskip-#2pt%upper horizontal
\rule{#2pt}{#1pt}}% right vertical

\newcommand{\Yfund}{\raisebox{-.5pt}{\drawsquare{6.5}{0.4}}}%  fund
\newcommand{\Ysymm}{\Yfund\hskip-0.4pt%
                    \Yfund}%  symmetric second rank
\def\symm{\Ysymm}
\def\bsymm{\overline{\Ysymm}}

\def\drawbox#1#2{\hrule height#2pt
        \hbox{\vrule width#2pt height#1pt \kern#1pt
              \vrule width#2pt}
              \hrule height#2pt}

\def\Fund#1#2{\vcenter{\vbox{\drawbox{#1}{#2}}}}
\def\Asym#1#2{\vcenter{\vbox{\drawbox{#1}{#2}
              \kern-#2pt       % line up boxes
              \drawbox{#1}{#2}}}}

\def\fund{\Fund{6.4}{0.3}}
\def\asymm{\Asym{6.4}{0.3}}
\def\bfund{\overline{\fund}}
\def\basymm{\overline{\asymm}}

\def\Acknowledgements{\bigskip  \bigskip {\begin{center} \begin{large}
             \bf ACKNOWLEDGEMENTS \end{large}\end{center}}}

\begin{document}\begin{titlepage}
\rightline{{CPTH-S757.1299}}
\rightline{{LPTENS 99/55}}
\vskip 1cm
\centerline{{\Large \bf Non-Tachyonic Type 0B Orientifolds,}}
\vskip 0.2cm
\centerline{{\Large \bf Non-Supersymmetric Gauge Theories}}
\vskip 0.2cm
\centerline{{\Large \bf and Cosmological RG Flow}}
\vskip 1cm
\centerline{{\bf Carlo Angelantonj}${}^\dagger$ and 
{\bf Adi Armoni}${}^\ddagger$}
\vskip 0.5cm
\centerline{${}^\dagger$ Laboratoire de Physique Th{\'e}orique de
l'{\'E}cole Normale Sup{\'e}rieure\footnote{Unit{\'e} mixte du CNRS et de
l'ENS, UMR 8549}}
\centerline{24, rue Lhomond, F-75231 Paris Cedex 05}
\centerline{\bf angelant@lpt.ens.fr}
\vskip 0.3cm 
\centerline{${}^\ddagger$ Centre de Physique Th{\'e}orique de l'{\'E}cole 
Polytechnique\footnote{Unit{\'e} mixte du CNRS et de l'EP, UMR 7644}}
\centerline{F-91128 Palaiseau Cedex}
\centerline{\bf armoni@cpht.polytechnique.fr}
\begin{abstract}
We discuss gauge theories on D3 branes embedded in special 
non-tachyonic orientifolds of the 0B string theory.
In general, they correspond to non-supersymmetric $SU(N)$ gauge theories
with scalars in the adjoint representation and spinors in the 
(anti-)symmetric representation. We study these theories via the AdS/CFT
correspondence and present evidence of their relation to 
\Nfour SYM in the planar limit. We also discuss finite $N$ properties, 
focusing in particular on the renormalization group flow.
Up to two loops, the logarithmic running of the gauge coupling is
 described by the dilaton tadpole and the cosmological constant
that naturally emerge on the string theory side.
\end{abstract}
\end{titlepage}

\section{Introduction}

The study of the strong coupling regime of gauge theories is
 one the most difficult problems in high energy physics.
In recent years there was much progress in this direction
- mostly in the study of supersymmetric theories. Part of this
knowledge was gained with the help of string theory.

Lately, it was conjectured that the natural description of strongly
 coupled large $N$ \Nfour SYM 
theory is in terms of classical supergravity. This
 is known as the AdS/CFT correspondence \cite{ads} (for a recent
 review and references see \cite{review}).
 Although this
 conjecture is very powerful, its original version is restricted to
 theories with maximal supersymmetry. It seems very difficult to study
 non-supersymmetric and non-conformal gauge theories in this framework,
since it is not known how to construct their dual gravity backgrounds.

The first step in the direction of studying non-supersymmetric 
 conformal theories using the AdS/CFT duality was made in \cite{KS},
where
it was shown that one can break supersymmetry completely, still
within a large $N$ CFT. The idea was to study the type IIB string on
$AdS_5\times X_5$, where $X_5$ is a five-dimensional space
of the type
 ${S^5 / \Gamma}$. This construction yields conformal 
theories with a reduced
 number of supersymmetries or no supersymmetry at all. The conformal
 invariance of the resulting large $N$ theories was then proved both 
by string
 theory \cite{BKV,LNV} and by direct field theory techniques \cite{BJ}.  

A similar idea motivated
 the study of type 0B string theory\cite{KT1},  a
$(-1)^{F_s}$ orbifold of type IIB. The orbifold breaks
supersymmetry and the resulting string theory, purely bosonic, 
 contains the NS-NS sector of the IIB theory and doubled sets of R-R
 fields. In addition, it contains a tachyon from the twisted
 sector, while the NS-R and R-NS fermionic sectors are projected out.
More recently this theory was also related to an
orbifold of M-theory\cite{BG}.  

The theory that lives on the untwisted
(``dyonic'') D3 branes is an $SU(N)\times SU(N)$ gauge theory with
 6 adjoint scalars for each of the gauge groups and 4 $(\fund,\bfund)$
 + 4 $(\bfund,\fund)$ Weyl spinors.  
Recently, the construction
of \cite{BG} was used to study this gauge theory \cite{BCL}, that is 
conformal at large $N$ \cite{KT2,NS}.
Other type 0 conformal models were studied in \cite{BCR,AK,BFL2}.

Although this gauge theory seems to be perfectly sensible, one might
 suspect that, since the dual string theory contains a tachyon,
 some instabilities should appear on the gauge theory side,
and indeed, using the AdS/CFT
correspondence, it was shown 
that at strong coupling the dimension of the twisted operator
$T \sim {\rm tr} \ F_1^2 - {\rm tr} \ F_2^2$ 
becomes complex \cite{klebanov}. The
 presence of the tachyon makes the study of the dual gravity
 difficult, since the detailed structure of its coupling to the
 R-R fields is not known. Another (somewhat) disturbing feature is the doubling
 of the gauge factors in all models that live on ``dyonic"
D-branes \cite{AK}. This is due to the doubling of R-R fields,
and is seemingly unrelated to the tachyon issue.

Is there any way to project out the tachyon from the type 0 spectrum,
so that one has a stable dual gauge theory?

The construction of tachyon-free models was first discussed in 
\cite{sagnotti2}, 
where sign ambiguities in the Klein bottle amplitude \cite{pss} were used to
generalize the standard $\Omega$ projection. The non-tachyonic
projection required
a $U(32)$ gauge group on the D9-brane, together with chiral fermions in
the $\asymm$ and $\basymm$ representations. An exhaustive analysis of 
tachyon-free orbifold 
compactifications was initiated in 
\cite{carlo1}. In general, for the (would be) 
supersymmetric orbifolds new tachyons appear in the twisted
sector of 
the closed string theory and, for a geometric action of the orbifold
group, can not
be projected out by any (generalized) $\Omega$ projection. The $T^4 /Z_2$ and 
$T^6 / Z_3$ cases, however, are exceptional, since no tachyons appear in the
 twisted sectors
and thus one can find appropriate $\Omega$ projections that lead to
 non-tachyonic
open descendants. The analysis was then extended to $Z_N \times Z_M$ 
orbifolds \cite{kristin} and to the non-supersymmetric $T^6 /Z_2$ one 
\cite{alok}. A geometric interpretation of the signs in the 
Klein bottle amplitude, corresponding to the generalized $\Omega$ projections, 
has been given in \cite{BFL1}: the non-tachyonic projection involves the 
combined
action of the world-sheet parity with the right world-sheet fermion 
number, $(-1)^{f_R}$. 

In the simplest version of the construction, the orientifold of 0B
leads to a string theory that is very similar to the bosonic
part of type IIB \cite{sagnotti2}. From now on,
we shall refer to it as the type $0'$ string theory. 
The theory on the D3 branes close to O$'3$ planes of this sort was first
studied in \cite{BFL2}. It is an $SU(N)$ gauge theory with the field
content in table 1.
\begin{table}[H]
\begin{displaymath}
\begin{array}{l c@{ } c}
& \multicolumn{1}{c@{}}{SU(N)} \\
\hline
\rm {vector}  & adj.  \\
\rm {6\ scalars} & adj.  \\
\rm {4\ fermions} & \asymm + \basymm  \\
\end{array}
\end{displaymath}
\caption{The field content of the non-supersymmetric gauge theory
 on $N$ D3 branes close to O$'3$ planes of 0B string theory.}
\label{theory}
\end{table}

 Most of the present paper is devoted
to the study of this gauge theory and its gravity dual. We will show
that the theory is conformal at large $N$, using both string theory
and field theory arguments. 
 Since the bulk
theory does not contain any tachyon, the gauge theory 
should be stable. Moreover, its study via gravity will be rather
simple, and we will find an $AdS_5 \times S^5$ solution of the tree level
gravity action.

We are also interested in finite $N$ non-supersymmetric theories.
In particular, we consider D3 branes of non-tachyonic orientifolds of type
0B string theory, {\it i.e.} D3 branes in a background of O$'7$ planes and
32 D7 branes (together with their images). At finite $N$, the
theory is no longer conformal. On the string theory side, the
breakdown of conformal invariance is due to 
a dilaton tadpole and to a
cosmological constant that appear once $g_{st}$ is turned on.  
We use these stringy corrections
 to exhibit a qualitative logarithmic running of the coupling
constant, up to two loops on the field theory side.
The dilaton tadpole affects the one loop  beta function, 
while the cosmological
constant affects the two loop beta function. We also comment on the relation
of the dilaton equation of motion to the all orders beta function equation.

The organization of the paper is as follows.
In section 2 we describe the orientifold projections allowed in 0B 
models. In particular, we focus on the projections that remove
the tachyon from the 0B string spectrum, and  
construct the gauge theory on two different systems: 
O$'3$-D3 in type 0B and D3 branes on type $0'$ backgrounds.
We also discuss more involved models originating from D3 branes at
orbifold singularities, that lead to other non-supersymmetric field 
theories on non-tachyonic string backgrounds. Sections 3 and 4 are
devoted to the study of the gauge theories introduced in section 2 via
the AdS/CFT correspondence. We also study the field theory 
renormalization group flow via gravity (see \cite{GPPZ,RG,BVV,VV} for a recent
discussion and \cite{klebanov2} for another example of RG flow). In
section 5 we give an  independent argument for the conformal invariance of the
large $N$ theory, based on the work of \cite{kakushadze}. In section 6 we study
the relation between \Nfour SYM and the theory on the D3 branes. We suggest
that these theories become equivalent in the planar limit. In section
7 we discuss briefly other field theories constructed in
section 2.  

\section{Type 0 orientifolds}

In this section we construct some brane configurations that will be
further analyzed in the following sections. In particular, we
construct gauge theories corresponding to D3 branes embedded in 
non-tachyonic 0B orientifolds and O$'3$-D3 systems in 0B theories, where
the orientifold planes involve the non-tachyonic projection of 
\cite{sagnotti2}.
To this end, let us recall some known facts about the type 0B
string and its orientifolds. The torus partition function 
\beq
{\cal T} = \int_{\cal F} {d^2 \tau \over \tau_{2}^{6}} \ {1\over |\eta|^{16}}
\left\{ |O_8|^2 + |V_8 |^2 + |S_8 |^2 + |C_8 |^2 \right\}
\label{torus}
\eeq
can be conveniently written in terms of level-one $SO(8)$ characters,
whose expressions in terms of Jacobi theta functions are
\bea
O_8 &=& {1 \over 2\eta^4} (\vartheta^{4}_{3} + \vartheta_{4}^{4} )\ ,
\qquad V_8 = {1\over 2\eta^4} (\vartheta^{4}_{3} - \vartheta^{4}_{4}
) \ ,
\nonumber
\\
S_8 &=& {1 \over 2\eta^4} (\vartheta^{4}_{2} - \vartheta^{4}_{1} ) \ ,
\qquad C_8 = {1\over 2\eta^4} (\vartheta^{4}_{2} + \vartheta^{4}_{1})
\ .
\nonumber
\eea
From ${\cal T}$ one can easily extract the low-lying excitations, that
comprise a tachyon, the graviton, the dilaton and the Kalb-Ramond
field from the NS-NS sectors, and two scalars, two 2-forms and one
4-form from the R-R sectors. The doubling of the
R-R sector with respect to type IIB 
implies that the 0B string has two types of D$p$-branes for
odd $p$.

To this left-right symmetric theory one can associate orientifolds
\cite{cargese,sagnotti1}, that in general contain two different types of
orientifold planes and D-branes, since the R-R sector is now
doubled. A generic feature of the standard 0B orientifolds is the
presence of the tachyon in the unoriented closed sector. Actually,
using the properties of the underlying two-dimensional conformal field
theory \cite{pss}, it was shown in \cite{sagnotti2} that 
the 0B string admits three
inequivalent orientifold projections. These are related to additional
(world-sheet and space-time) symmetries,
and, in particular, the combination $\Omega (-1)^{f_R}$, with
$\Omega$ the world-sheet parity and $f_R$ the world-sheet fermion
number, removes the tachyon from the closed
unoriented spectrum \cite{sagnotti2}, 
thus stabilizing the vacuum. Moreover, the $\Omega (-1)^{f_R}$
projection keeps only one copy of the  R-R sector, thus allowing for
the presence of only one type of D$p$-branes for any odd $p$ (as in the
IIB theory). 
An exhaustive analysis of orbifold compactifications has shown that one
can remove the closed tachyon only in the $T^4 / Z_2$ and
$T^6/Z_3$ \cite{carlo1}, in the $T^6/Z_2 \times Z_2$ \cite{kristin} and
in the non-supersymmetric $T^6 /Z_2$ \cite{alok} orbifolds.

\subsection{The O$'3$-D3 system}

Let us describe the simplest model on D3 branes at ``non-tachyonic''
orientifold planes.
The O$'3$ planes  project the D3 brane gauge field by $\Omega ' =
\Omega (-1)^{f_R} {\cal I}_6$, where  ${\cal I}_6$
inverts the six (non-compact) coordinates transverse to the D3-brane.

On a flat Minkowski background, the annulus and M{\"o}bius
amplitudes compatible with the type 0B string are
\bea
{\cal A} &=& {1\over 2} \int_{0}^{\infty} {dt \over t^3} {1\over
\eta^8} \ \left( 2 N \bar N \, V_8 - (N^2 + \bar N ^2 ) \, S_8 \right)\ , 
\label{d3brane}
\\
{\cal M} &=& \pm {1\over 2} \int_{0}^{\infty} {dt\over t^3} {1\over
\hat\eta^8} \ (N+\bar N) \ \hat S _8  \ ,
\nonumber
\eea
where the $\eta$-function and the $SO(8)$ characters depend on the
modulus of the doubly-covering torus ($it/2$ for the
annulus and $(1+it)/2$ for the M{\"o}bius strip). 
One can simply extract the massless fields that live on the 
O$'3$-D3 system: for $N$ branes
they are gauge bosons and six scalars in the adjoint of a $U(N)$ gauge
group, as well as four Weyl spinors in the $\asymm + \basymm$ 
representations. Actually, since the space transverse to the D3 branes
is non-compact, the R-R tadpole need not be imposed, and 
one has the freedom to revert the sign in the M{\"o}bius 
amplitude thus obtaining spinors in symmetric representations.

The $\Omega '$ projection acts also on the bulk theory. In fact,
as we will see in the next section,
in the spirit of the AdS/CFT correspondence, 
gauge invariant operators on D3
branes are related to harmonics of the bulk fields on $AdS_5 \times X_5$, with
$X_5$ some appropriate Einstein manifold respecting the symmetry of the brane
configuration.
In our case $X_5 = S^5/Z_2 = RP^5$, where the freely-acting
$Z_2$ corresponds to the $\Omega '$ projection, and thus only the
harmonics of the projected (bulk) fields survive
\cite{witten2,review}. 

Aside from the massless excitations, one can extract other interesting
informations from the amplitudes \eqref{torus} and
\eqref{d3brane}. For instance, since the
fundamental ($N$) and anti-fundamental ($\bar N$) representations of
unitary groups have identical dimensions, 
one can easily show that the annulus amplitude vanishes, as 
\beq
{\cal A} \sim N^2 \ (V_8 - S_8) \sim N^2\ \left( \vartheta_{3}^{4} -
\vartheta_{4}^{4} - \vartheta_{2}^{4} \right) \equiv 0 \ .
\eeq
This translates into a vanishing force between pairs of D3 branes, as a result
of a perfect cancellation between the gravitational (NS-NS) attraction
and the R-R repulsion. 

Since gauge/string theory is non-supersymmetric, one would expect the presence
of a dilaton tadpole and/or of a cosmological 
constant on the string theory side.
Unfortunately, it is hard to perform explicit string theory
computations in curved backgrounds and therefore, in the following we will
present explicit models where these corrections are under control and a precise
comparison between string theory and gauge theory can be worked out even at
finite $N$.

\subsection{D3 branes in type $0'$ backgrounds}

The configuration that now we are going to construct corresponds to
D3 branes in non-tachyonic orientifolds of the type 
0B string. In particular the background is an eight-dimensional toroidal 
compactification of the $\Omega'$ projected type 0B. Before discussing the 
gauge theory arising on the D3 branes, let us introduce the background string
theory and discuss its main features. 

The starting point is the halved toroidal partition function of 
the 0B string on a 
two-dimensional Narain lattice with momenta
\beq
p_{L,R}^{i} = {m_i \over R} \pm {n_i R \over \alpha '} \ .
\eeq
The non-tachyonic Klein bottle projection (associated with $\Omega (-1)^{f_R}
{\cal I}_2$),
\beq
{\cal K} = {1\over 2} \int_{0}^{\infty} {dt \over t^5} \ {1\over \eta^8}\ 
\left( - O_8 + V_8 - S_8 + C_8 \right) \sum_{n} e^{-\pi t R^2 n^2 / \alpha '}
\ ,
\eeq
leads to the following massless spectrum: the eight-dimensional metric
tensor, the moduli of the internal $T^2$, the dilaton, two abelian vectors
related to the mixed components of the NS-NS $B$-field, and full
ten-dimensional 0-form, 2-form and self-dual 4-form R-R potentials.

The open sector that completes the orientifold construction \cite{cargese} is
encoded in the annulus and M\"obius amplitudes
\bea
{\cal A} &=& {1\over 2} \int_{0}^{\infty} {dt \over t^5} \ {1\over \eta^8}
\ \sum_{j,k=1}^{4} \sum_{n} \left[ V_8 \ \left( M_j \bar M_k \ e^{-\pi t R^2
(n + a_j - a_k)^2 /\alpha '} \right. \right.
\nonumber 
\\
&& + \left. \bar M_j M_k \ 
e^{-\pi t R^2 (n - a_j + a_k )^2 /\alpha '} \right) \
-
\ S_8\ \left( M_j M_k \ e^{-\pi t R^2 (n + a_j + a_k )^2 /\alpha '} \right.
\nonumber
\\
&& \left. \left. + \bar M _j 
\bar M_k \ e^{-\pi t R^2 (n - a_j - a_k )^2 / \alpha '}
\right) \right] \ , \nonumber
\\
{\cal M} &=& {1\over 2} \int_{0}^{\infty} {dt \over t^5} \ {1\over \eta^8}\ 
\hat S _8 \ \sum_{j=1}^{4} \sum_n \left( M_j \ e^{-\pi t R^2 (n + 2 a_j )^2
/\alpha '} + \bar M _j \ e^{-\pi t R^2 (n - 2a_j )^2 /\alpha '}
\right) \ , \nonumber
\eea
where the $M_j$ and $\bar M _j$ describe Chan-Paton multiplicities.
Here we have decided to equally distribute 
the D7 branes on the four O$'7$ planes 
\beq
a_1 = (0,0)\ , \quad a_2 = (1/2,0)\ , \quad a_3 = (0,1/2)\ , \quad 
a_4 = (1/2 ,1/2) \ ,
\eeq
in order to cancel tadpoles locally. 
After imposing (R-R) tadpole cancellation the gauge theory on 
each set of D7 branes has a $U(8)$ gauge group with 2 scalars in the
adjoint and two spinors in the $\asymm$ and $\basymm$. 

Actually, for $\Omega'$ orientifolds one is always left with a non-vanishing 
dilaton tadpole coming from the ($V_8$ character in the) annulus
amplitude, as can be appreciated from the massless contributions to the
transverse amplitudes:
\bea
\tilde{\cal K} &\sim& - {2^6 \over 2} S_8 \ , \nonumber
\\
\tilde{\cal A} &\sim& {2^{-6} \over 2} \left[ \left( \sum_{j=1}^{4} M_j + \bar
M_j \right)^2 (V_8 - S_8) - \left( \sum_{j=1}^{4} M_j - \bar M_j \right)^2
(O_8 - C_8 ) \right] \ , \nonumber
\\
\tilde{\cal M} &\sim& {2\over 2} \left( \sum_{j=1}^{4} M_j + \bar M_j \right)
\hat S_8\ . \nonumber
\eea
Taking into account R-R tadpole conditions the numerical value of the 
dilaton tadpole is then
\beq
C^2 = 2^{-6}\, \left( \sum_{j=1}^{4} M_j +\bar M _j \right)^2 = 
2^{-6}\, 2^{12} = 8^2 \ . \label{tadpole}
\nonumber
\eeq
This has to be contrasted with the supersymmetric case, where the 
dilaton couples both to D-branes and O-planes and 
its tadpole vanishes as a result of R-R tadpole conditions.
Moreover, the torus, Klein bottle and M\"obius amplitudes are non-vanishing,
and this translates into the emergence of a one-loop cosmological constant. 
Thus, dilaton tadpole and cosmological constant induce for the dilaton field
a potential of the form
\beq
V (\Phi ) = C e^{-\Phi} + 2 \Lambda\ ,
\eeq
whose properties will be further analyzed in section 4.

We are now ready to add $N$ D3 branes in the background of this type
$0'$ string. The reduced $SO(8)\to SO(4)\times SO(4)$ transverse Lorentz
symmetry translates into a breaking of the $SO(8)$ characters into products of
$SO(4)$ ones:
\bea
O_8 &= O_4 O_4 + V_4 V_4 \ , \qquad S_8 &= S_4 S_4 + C_4 C_4\ ,
\\
V_8 &= V_4 O_4 + O_4 V_4 \ , \qquad C_8 &= S_4 C_4 + C_4 S_4\ .
\eea
Putting all the D3 branes close to the orientifold plane at
the origin ($a_1 = (0,0)$) results in the following (massless) deformation
of the annulus and M\"obius amplitudes:
\bea
{\cal A} &\sim& (M_1 \bar M_1 + N\bar N) (V_4 O_4 + O_4 V_4) 
\nonumber 
\\
&&- {1\over 2}\
(M_{1}^{2} + \bar M_{1}^{2} + N^2 + \bar N^2 ) (C_4 C_4 + S_4 S_4 )
\nonumber
\\
&& + (\bar M_1 N + M_1 \bar N) \ C_4 O_4 - ( M_1 N + \bar M _1 \bar N ) \ C_4
O_4 \ ,
\\
{\cal M} &\sim& (M_1 + \bar M_1 + N + \bar N )\ C_4 C_4 +
(M_1 + \bar M _1 - N - \bar N )\ S_4 S_4 \ .
\eea
The field theory on the D3 branes is thus an $SU(N)$ gauge theory with 
the charged matter in table 2. (We are neglecting the $U(1)$ factor 
associated with the center of mass of the brane configuration.)

\begin{table}[H]
\begin{displaymath}
\begin{array}{l c@{ } c}
& \multicolumn{1}{c@{}}{SU(8) \times SU(N)} \\
\hline
\rm {vector} & (adj,1) \\
\rm {6\ scalars} & (adj,1) \\
\rm {4\ Weyl\ spinors} & (\asymm + \basymm ,1) \\
\hline
\rm {vector}  & (1,adj)  \\
\rm {6\ scalars} & (1,adj)  \\
\rm {2\ Weyl\ spinors} & (1,\asymm + \basymm)  \\
\rm {2\ Weyl\ spinors} & (1,\symm + \bsymm)  \\
\hline
\rm {2\ scalars} & (8,\bfund ) + (\bar 8, \fund ) \\
\rm {1\ Weyl\ spinor} & (8,\fund) + (\bar 8 ,\bfund) \\
\end{array}
\end{displaymath}
\caption{The field content of the non-supersymmetric gauge theory
 on $N$ D3 branes in type $0'$ string theory.}
\label{theory2}
\end{table}

\subsection{D3 branes at orbifold singularities}

We can now turn to more complicated models, like O$'3$-D3 systems on orbifold 
singularities. The analysis of \cite{carlo1} 
suggests to restrict oneself to the 
$C^2/Z_2$ and $C^3 /Z_3$ singularities, since the other cases would introduce 
additional (twisted) tachyons in the bulk theory. For supersymmetric strings,
$Z_2$ and $Z_3$ singularities would correspond to \Ntwo and \None
gauge theories
\cite{KS}, and in 
our case one finds an $SU(N) \times SU(N)$ gauge group for O$'3$-D3 
systems on a $C^2/Z_2$ singularity. The charged matter comprises four
 scalars in the
$(\fund , \fund )$ and $(\bfund , \bfund )$ and Weyl spinors in
 the $(\fund, \bfund )$, 
$(\bfund , \fund )$, $(\asymm + \basymm ,1)$ and $(1, \asymm + \basymm )$ 
representations. 

The $C^3 /Z_3$ singularities lead to a chiral spectrum containing
 an $SU(N-4) \times 
SU(N) \times SU(N)$ gauge group, as well as Weyl spinors in
 the $(\asymm + \basymm, 1)$,
$(1, \fund , \fund)$, $(1, \bfund , \bfund)$
 representations. Moreover, there are
three copies of Weyl spinors in the $(\fund , \fund ,1$), $(\bfund , 1 ,
\bfund )$, $(1, \basymm ,1)$ and $(1,1,\asymm)$ representations and
 three copies 
of complex scalars in the $(\bfund , \fund ,1)$, $(\fund ,1,\fund )$, 
$(1, \fund , \fund )$ and $(1, \bfund , \bfund )$ representations.

All these models have the same main features of the O$'3$-D3 system in
 flat Minkowski
space-time: there is no brane-brane interaction, 
and one expects that a (half-loop) dilaton tadpole and a 
(one-loop) cosmological constant be
 generated.

\section{The low energy action and the AdS/CFT correspondence}

The orientifold projection introduced in the previous
section leads to a non-supersymmetric string theory with a
rather simple low energy action. The action of
$\Omega '= \Omega (-1)^{f_R}$ on the
 oscillators of the underlying CFT is \cite{BFL1}
\bea
 & & \Omega ' \ \alpha ^\mu _n \ \Omega ' = \tilde \alpha ^\mu _n \\
 & & \Omega ' \ \Psi  ^\mu _r  \ \Omega '   = \tilde \Psi ^\mu _r \\
 & & \Omega ' \ \tilde \Psi ^\mu _r \ \Omega '   = \Psi ^\mu _r \\
 & & \Omega ' \ \state{0} _ {NS-NS} = - \state{0} _{NS-NS}
\eea
Therefore, whereas type 0B string theory contains
 a tachyon and two sets of R-R fields, the orientifold projection removes
 the tachyon and one set of R-R fields from the bulk. It picks the 
R-R two-form from the $(R+,R+)$ sector and the zero-form and four-form
 from the $(R-,R-)$ sector. Thus the
 resulting ten-dimensional $0'$ theory 
is similar to the bosonic part of type IIB
\beq
S_0 ^{(\Omega)} = 
\int d^{10} x \sqrt {-G}
\left ( e^{-2\Phi} (R+4 \partial _\mu \Phi \partial ^\mu \Phi )  
-{1\over 2} \left(|F_1| ^2 + |F_3| ^2  +
 {1\over 2} |F_5|^2 \right)\right ) 
\label{treeaction}
\eeq

Note that the tachyon is projected out and hence we expect the
dual theory on the D3 branes to be more stable than for other type 0
models. We will make this statement more precise in the following. 
Note also that the NS-NS 2-form $B_{\mu \nu}$ is projected out.
A final remark
concerning the action \eqref{treeaction} is
that, in contrast to the unprojected type 0 action, we have only one set of R-R
fields. The doubling of the R-R fields in the type 0B action leads
to a semi-simple $SU(N)\times SU(N)$ gauge theory on the dyonic D3 branes.
Here we expect that the dual theory on the brane have only
one gauge factor. Indeed, the large $N$ theory has a simple $SU(N)$ 
gauge group.

In order to study the type $0'$ theory in the presence of D3 branes
we first perform 2 T-dualities along the $8,9$ directions and than
place $N$ D3 branes. The resulting background consists of O$'7$ and
8 D7 branes and localized $N$ D3 branes. (Here we are interested in the
degrees of freedom living close to the orientifold plane at the
origin.) The near horizon background, 
$AdS_5 \times S^3$ (where the $S^3$ sits inside the $S^5$) 
respects the $SU(2)$ flavor
symmetry of the model (see table 2). At infinite $N$ the 8 D7 branes can
be neglected and the flavor symmetry is enhanced to $SU(4)$. Accordingly, 
in this limit one finds the standard $AdS_5 \times S^5$ background metric
\beq
ds^2 = d\tau ^ 2 + e^{2\tau} dx_i ^2 + d\Omega _5 ^2 \label{metric}
\eeq
(with a radius $R=1$).
The $AdS_5$ part of the metric
corresponds to an $SO(2,4)$ symmetry and suggests that the
large $N$ non-supersymmetric dual theory is conformal. The $S^5$ part
corresponds, in the present case, to an $SU(4)$ global flavor symmetry.

The situation for the O$'3$-D3 system is apparently quite different. 
In this case the bulk theory is the full 0B string and 
(in the large $N$ limit where the charge of the orientifold 3-plane
can be neglected and the configuration is indeed extremal) the near 
horizon geometry is $AdS_5 \times RP^5$. However, the $Z_2$ 
identification in $RP^5$ keeps only the invariant harmonics of the
bulk fields, and the resulting spectrum coincides with the 
($\Omega '$) projected bulk theory.
Further evidence for the conformal invariance of these models will 
be given in sections 5 and 6.

The AdS/CFT correspondence can be used to study the stability of 
these non-supersymmetric models\cite{klebanov}.
 In the type 0 (unprojected) theory, the
bulk theory contains a tachyon. The squared-mass of the tachyon is shifted
to a positive value at small radius of the $AdS$ metric \cite{KT2}. However, at
large values of the radius, when the bulk theory is flat, we really
have a tachyon. It is related, by holography, to the twisted
operator
\beq
T \sim {\rm tr} \ F_1 ^2 - {\rm tr} \ F_2 ^2 \ ,
\eeq
and therefore it was argued \cite{klebanov} that at large 't Hooft
 coupling $\lambda$ (large
$AdS$ radius) the dimension of ${\rm tr} \ F_1 ^2 - {\rm tr} 
\ F_2 ^2$ will become
complex, $\Delta(\lambda) =2 + \sqrt{4-2\sqrt {2\lambda}}$. 
This indicates an instability in the gauge theory at strong coupling.

The present bulk theory contains no tachyon. Therefore, the
corresponding gauge theory should be stable even at strong
coupling. This conclusion is not surprising since, as we shall see in
section 6, this theory is related to \Nfour SYM.

\section{Cosmological renormalization group flow}

In the previous section we started a study of the gauge theory
using the bulk theory. The discussion was restricted to large $N$,
where the gauge theory is expected to be conformal.

Here we would like to study ${1\over N}$ corrections to
the gauge coupling of the theory on the D3 branes in the type $0'$ 
background (the field content is described in table 2). 
Our model is the non-supersymmetric analogue of the one studied
in\cite{APTY,GNS}.
At finite $N$ the coupling is expected to run, and indeed the value
of the one-loop beta-function coefficient is 
\beq
 b_1 = -8. \label{b1}
\eeq

In calculating \eqref{b1} we took into account the contribution of the
8 flavors of fermions and scalars (arising from the open strings
that connect the D3-D7 branes).
Note that the value \eqref{b1} is ${1\over N}$ suppressed with 
respect to the
general expected behavior of beta function coefficients and that the
theory is IR free. The 
higher order beta function coefficients are also expected to behave
as 
\beq
b_n \sim N^{n-1}\ , \label {bn}
\eeq
as befits with the conformal invariance of the model at large $N$.

The beta function equation (up to two loops) reads
\beq
{d g \over d\log u } = - { g^3 \over {(4\pi)}^2} b_1 - {g^5 \over
  {(4\pi)}^4} b_2 + ... \ \ ,
\eeq
and its solution is
\beq
g^2 (u) = {8\pi^2 \over b_1 ({\log {u\over u_0} + {b_2 \over 2b_1^2}\log
    \log {u\over u_0}})} + ... \ \ .\label{gaugeres}
\eeq
 
In order to study these ${1\over N}$ corrections on the gauge theory side,
we should go beyond the tree-level classical gravity on the string
theory side. Before entering into details, we would like to mention
that several attempts to study the running of the coupling were already made
in various type 0 models \cite{log}. Thus, the coupling of the tachyon
to the dilaton was used to extract the logarithmic 
 running of the gauge coupling in several non conformal models,
but it was difficult to find precise results,
since the detailed structure of the tachyon 
coupling to the dilaton is not known. Moreover, these results depend 
on the minimum of the tachyon potential (the function that describes
the dilaton-tachyon couplings), and it is not clear whether such a
minimum really exists.

In the present case, the spectrum does not contain 
a tachyon, and therefore the
mechanism that we present seems more controlled. An important remark
is that at finite $N$ the background will no longer be an $AdS_5\times
S^3$, due to the 
back-reaction of the dilaton on the metric. However, we shall see
that the leading ${1\over N}$ correction is determined by the tree level
$AdS_5\times S^3$ background. 

When $g_{st}$ is
turned on, the classical action \eqref{treeaction} acquires
corrections, and we find two important contributions: a cosmological
constant due to the loss of supersymmetry
and a dilaton tadpole arising from the annulus diagram (see section
2). Generally, the
cosmological constant is expected to be proportional to $N$ (the number
of branes), but is also suppressed by
$g_{st} ^2 \sim {1\over N^2}$ with respect to
the tree level action. The dilaton tadpole is suppressed
by $g_{st} \sim {1\over N}$. The corrected action is
\beq
S=S_0 ^{(\Omega)} + \int d^{8} x \sqrt {-G} (C e ^{-\Phi} + 2\Lambda) 
\ , \label{action} 
\eeq    
and the eight-dimensional equations of motion in the string frame are
\bea
e ^{-2\Phi} \left ( 8 {(\nabla \Phi )}^2 - 8 \nabla ^2 \Phi -2R
 \right ) - Ce ^{-\Phi} &=& 0 \ ,
\label{motion1} 
\\
e ^{-2\Phi} \left ( R_{\mu \nu} + 2 \nabla _\mu \nabla _\nu \Phi
 \right ) - {1\over 2} G_{\mu\nu} \left( {C\over 2} e^{-\Phi} + 2\Lambda 
\right) + f_{\mu\nu} ({\rm RR}) &=& 0 \ , \label{motion2}
\eea
where the traceless tensor $f_{\mu\nu}$ depends on the R-R fields, and 
in particular on the R-R 5-form field strength. Substituting the
trace of \eqref{motion2} in \eqref{motion1}, we obtain for the dilaton
the following equation: 
\beq
\nabla ^ 2 e ^{-2\Phi} = {5 C\over 2}  e^{-\Phi} + 8 \Lambda \ . 
\label{dilaton}
\eeq
Clearly, $\Phi = \Phi _0$ is no longer a solution.
In principle, one should also study the equation of the metric.
However, as an approximation, we will assume that the background
is still $AdS_5 \times S^3$. 
This assumption is valid (in the string frame,
where the metric does not depend on the dilaton) because the left hand
side of \eqref{dilaton} is dominated by the tree level background. 
The back-reaction of the dilaton on the metric
will be of higher order in ${1\over N}$. Assuming $\Phi =\Phi(\tau)$
and substituting
\eqref{metric} in \eqref{dilaton}, we obtain
\beq
\partial _{\tau} ^2 e ^{-2 \Phi} + 4 \partial _{\tau} e ^{-2 \Phi} 
= {5C\over 2} e ^{-\Phi}  +8 \Lambda \ .
\eeq
The Ansatz
\beq
e ^{-\Phi} = A \tau + B \log \tau ,
\eeq
gives an approximate solution at large values of $\tau$, if
\bea
& & A = {5 C\over 16} \label{Aeq} \\
& & B =  {16 \Lambda \over 5 C} - {5 C \over 64} \label{Beq}.
\eea
Generally, the ${5 C\over 64}$ in \eqref{Beq} is
expected to be negligible with respect to ${16\Lambda \over 5 C}$ 
in the large $N$ limit, since $\Lambda \sim O(N)$.

Recall that the coordinate $\tau$ is related to the energy
scale $u$, through the standard $AdS_5$ metric
parameterization, by
\beq
\tau = \log {u\over u_0}\ ,
\eeq
while the dilaton is related to the gauge coupling by
\beq
g _{YM} ^2 \sim g_{st} = e ^{\Phi}\ .
\eeq
Thus we obtain
\beq
{1\over g_{YM} ^2} \sim A \log {u\over u_0} + B \log \log {u\over u_0} \ ,
\label{gravityres}
\eeq
with a nice qualitative agreement between gravity \eqref{gravityres} 
and field theory calculations \eqref{gaugeres} up to two loops. The
dilaton tadpole determines the one-loop beta function, while 
the cosmological constant determines the two-loop beta 
function.
Remarkably, in our model both the dilaton tadpole \eqref{tadpole} 
and $b_1$ \eqref{b1} 
are proportional to the number of D7 branes.
This result 
is encouraging, since it links the AdS/CFT correspondence to more
realistic models where conformal invariance and supersymmetry 
no longer exist.

 It might look surprising that we find an agreement
between gravity and perturbative Yang-Mills. After all, we do not
expect that dual descriptions in terms of gluons and gravitons 
apply at the same time.
The reason for the qualitative 
agreement is the following: As we shall see in section 
6, the non-supersymmetric models (table 1 and table 2) converge
 at infinite $N$ to \Nfour SYM. It is believed that in this case,
 the background is unaffected by $\alpha '$ corrections. In the present
 case, we add a perturbation of ${1\over N}$. Our hope is that for large
enough $N$ and fixed energy $u$, our results are unchanged. 

Finally, note that the dilaton
equation of motion reproduces qualitatively, at least to leading 
order in ${1\over N}$, all
orders of the running coupling equation. Generally, in the presence
of an orientifold, the string theory effective action should receive
corrections of any power of $g_{st}$ (even and {\em odd}), and
therefore the dilaton equation \eqref{dilaton}, should read
\beq
\nabla ^ 2 e ^{-2\Phi} = 
\sum _ {n=-1} ^\infty  C_n e ^ {n\Phi} \label{dilaton2}.
\eeq
With the identification $g_{st} \sim g^2$ and an
$AdS_5 \times X_5$ background, we find
\beq
\partial _{\tau} ^2 {1\over g^4} + 4 \partial _{\tau} {1\over g^4} 
= \sum _ {n=-1} ^\infty  C_n g ^{2n} \label{dilaton3} \ .
\eeq
The two derivative term (the first term) in \eqref{dilaton3} gives a
negligible contribution at large $\tau$ and large $N$, since the additional
derivative with respect to $\tau$ yields terms which are ${1\over \tau}$
with respect to one derivative contributions. 
An example is the 
${5 C\over 64}$ in eq.\eqref{Beq}, that comes from the two derivative
term. Upon neglecting the two derivative term, one obtains
\beq
{dg \over d\log u} = - \sum _{n=-1} ^\infty {C_n \over 16} g^{2n+5}\ ,
\eeq
the all orders beta function equation for the Yang-Mills
coupling !

\section{String theory amplitudes and the planar limit}

In this section we want to use the results of \cite{kakushadze} to 
give plausible arguments about the finiteness of the gauge theories
previously introduced. For simplicity we will focus on the
O$'3$-D3 system in the large $N$ limit, although similar arguments apply
also to the other model. 

In a theory with D-branes and orientifold planes string amplitudes involve 
world-sheets with $b$ boundaries, $c$ crosscaps and $h$ handles, 
and are weighted by
\beq
(N g_{st})^b \, g_{st}^{c} \, g_{st}^{2h-2} = 
\lambda ^{2h-2+b+c} \, N^{-c-2h+2} \ . \label{kak1}
\eeq
In particular, for the large $N$ behavior of $M$-point correlators
of fields charged under the gauge group there are two classes of 
diagrams that have to be considered: diagrams without handles and
crosscaps, and diagrams with handles and/or crosscaps, that from 
\eqref{kak1} are subleading in the large $N$ limit. 
In the same way, non-planar diagrams with only boundaries are suppressed
with respect to planar ones. 

For planar diagrams with $b$ boundaries and $M$ external lines, the boundary 
conditions must satisfy (in the notation of \cite{kakushadze})
\beq
\gamma^{\mu_1}_{a_1} = \prod_{s=2}^{b} \gamma_{a_s}^{\mu_s} \ ,
\eeq
where the $\gamma$'s define the action of a (generic) orbifold group on the
Chan-Paton matrices and $s=1$ refers to the outer boundary, while 
$s=2, \ldots ,b$ refer to the inner boundaries. 

If the orbifold group is trivial, in the large $N$ limit the 
$M$-point amplitude 
behaves as in the \Nfour one without orientifold planes (with gauge
group $U(N)$). For non-trivial (space-time) orbifold groups, it has
been shown in \cite{kakushadze} that diagrams with twisted Chan-Paton 
matrices are vanishing or suppressed in the large $N$ limit, thus 
leading to the conjecture that supersymmetric O$'3$-D3 systems on 
orbifold singularities have the same leading behavior as the \Nfour
$U(N)$ gauge theory. 

Actually, the model that we are considering can be thought of as an
orbifold of the supersymmetric O$'3$-D3 system, where the orbifold 
group corresponds to the internal $(-1)^{F_s}$ symmetry. This
projection leads to two different $\gamma$ matrices such that:
\beq
{\rm tr} (\gamma_1) = (N + \bar N) \ , \qquad 
{\rm tr} (\gamma_2) = i (N- \bar N ) \ .
\eeq
It is then evident how to apply the previous arguments to the present
case. Non-planar and unoriented amplitudes are suppressed in the
large $N$ limit, and among the planar diagrams, the ones with $\gamma_1$ 
boundary conditions on both external and internal boundaries give
the same behavior of \Nfour $U(N)$ gauge theory. On the other
hand, diagrams with at least one inner boundary with a $\gamma_2$ insertion
are vanishing since they are proportional to
\beq
\sum {\rm tr} (\gamma^{1}_{a_1} \, 
\lambda_1 \ldots \lambda_M ) \ \prod_{s=2}^{b}
{\rm tr} (\gamma^{\mu_s}_{a_s})
\eeq
and at least one of the inner $\gamma$'s is 
\beq
\gamma^{\mu_s}_{a_s} = \gamma_2 \quad \Rightarrow \quad
{\rm tr} (\gamma^{\mu_s}_{a_s}) = i (N-\bar N) =0 \ .
\eeq 

We thus conclude that in the large $N$ limit our model 
indeed behaves like
the \Nfour supersymmetric $U(N)$ gauge theory.
Similar arguments can also be applied to O$'3$-D3 systems of type 0B 
on orbifold singularities.

\section{Field-theory analysis - The relation to \Nfour SYM}

 Arguments in favor of conformal invariance can also be presented in a field theory context.
Our gauge theory corresponds to an $SU(N)$ gauge group with 6 scalars
in the adjoint representation and 4 spinors in the $\asymm$
representation and its conjugate.
The string theory analysis of the previous sections proved that this
theory is conformal in the large $N$ limit. In fact, as we shall see,
the large $N$ limit of this theory is exactly \Nfour SYM.

The similarity to \Nfour SYM is rather clear. The only difference between
 the two theories is that the spinors transform in a different
 representation, and therefore supersymmetry in broken explicitly in the
 type 0 case at finite $N$. However, we will argue that at large $N$
 adjoint spinors and spinors in $\asymm + \basymm$ give the same 
contribution to any amplitude.

Our arguments are similar to those given in \cite{BJ} for
the case of orbifold field theories, where
it was shown 
that the planar graphs of the \Nfour SYM and of the daughter
 theory, an orbifold truncation of \Nfour, are the same up to
a rescaling of the gauge coupling in the latter. 

Let us therefore 
examine the relation between the present non-supersym\-metric gauge
theory and the \Nfour theory. Since we are interested in the planar limit,
 it will be convenient to use 't Hooft's double index notation, where
fields that transform in the adjoint representation
carry two lines with arrows in opposite directions.
This is the case, since 
the adjoint representation is obtained by a multiplication
of fundamental and anti-fundamental representations. Similarly, a
field transforming in the anti-symmetric representation
carries two
lines with arrows in the {\em same direction}, since the
anti-symmetric representation is obtained by
 multiplication of two fundamentals. Thus the fermion propagator 
in the non-supersymmetric theory will be represented by two parallel
lines with arrows pointing in the same direction, while bosonic
propagators are
represented by two parallel lines with arrows pointing in opposite directions.

The Feynman rules of the non-supersymmetric theory and of the
supersymmetric theory are very similar. The only differences are in
the vertices which couple fermions and bosons, such as the coupling of the
 gluon to the fermions or the Yukawa coupling. In these graphs, one of
 the arrows of the fermion lines is reversed. The fermionic
 propagator and the fermion-boson coupling are described in figure 1 below. 

\begin{figure}[H]
  \begin{center}
\mbox{\kern-0.5cm
\epsfig{file=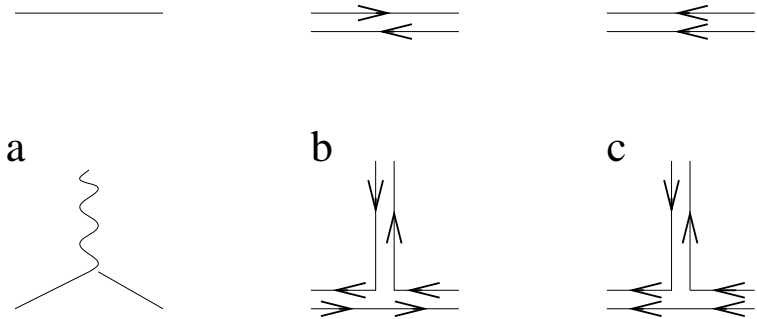,width=10.0true cm,angle=0}}
\label{rlues_fig}
  \end{center}
\caption{Feynman rules for \Nfour SYM and
  for the non-supersymmetric theory. a) Fermion propagator and
 fermion-boson vertex. b) Feynman rules for the supersymmetric theory: both
fermions and bosons are in the adjoint representation. c) In the 
non-supersymmetric orientifold theory bosons
 transform in the adjoint representation whereas fermions transform in
 the $\asymm$ representation (and in the $\basymm$ rep.).}
\end{figure}

Although we will not present a rigorous proof, we will argue that this change
of the Feynman rules will not affect any calculation that
only depend on planar diagrams: 
every planar diagram in \Nfour SYM can be deformed
so that it represents the non-supersymmetric 
theory and vice versa. As a result,
there is a one-to-one correspondence between the planar graphs
of both theories.

The rule is simple: one should follow the fermionic double lines and reverse 
 the arrow of one of the lines, so that the two arrows along the
 fermionic lines point in the same direction. It is easy to see
 that this rule is consistent with the Feynman rules,
and in particular the interaction
 of bosons and fermions is unaffected by this deformation: the bosons
 will remain in the adjoint representation.

Let us give a simple example. Consider the fermionic loop 
contribution to the one-loop vacuum
 polarization in both theories (figure 2). 
It is clear that the planar diagram
 of the supersymmetric theory can be deformed into the planar diagram of 
the non-supersymmetric theory.

\begin{figure}[H]
  \begin{center}
\mbox{\kern-0.5cm
\epsfig{file=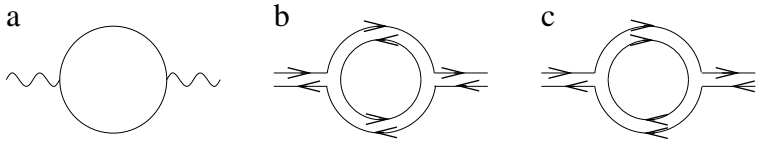,width=10.0true cm,angle=0}}
\label{vacuum_fig1}
  \end{center}
\caption{ a) The fermionic loop contribution to the vacuum
  polarization. b) A planar contribution in the supersymmetric 
theory. Fermions are in the adjoint representation. c) The arrows in
the internal loop were reversed such that the fermions are now in the 
$\asymm$ (and $\basymm$) representations.}
\end{figure}

One might be tempted to believe that all 
diagrams (not only the planar ones)
of the two theories are equivalent. This, however, is not true.
 There are non-planar diagrams that exists only in one of the
 theories. For instance, the one loop beta function is $b_1 = -{16\over 3}$
in the non-supersymmetric
 theory, but $b_1=0$ in the supersymmetric case.

The reason why the deformation works in the planar case is that
planar graphs form a collection of {\em non-intersecting} lines. One should
follow the fermionic line and simply reverse one of the arrows,
and  this
procedure does not lead to inconsistencies. The fermionic lines can be
closed loops, or lines with an end (in the case of external
legs). Since lines do not intersect and fermion number 
is conserved, the deformation can not change the representations of the
 bosons, that remain in the adjoint. 

The conclusion of this discussion is that in the infinite $N$ limit
 the non-supersymmetric theory becomes supersymmetric,
since there is no 
 difference between the anti-symmetric representation and the adjoint
 representation at large $N$. In this regime the theory described by the
 orientifold of type 0 becomes the \Nfour SYM.

Since this theory becomes \Nfour SYM in the planar
limit, we arrive at several
interesting conclusions. In this limit, the
 theory has BPS states, a moduli space of vacua and
possesses Olive-Montonen duality.
These features can be proved, in the infinite $N$ limit,
by the relation to \Nfour, via Feynman diagrams or via the bulk string
 theory. The existence of a moduli space of vacua can
  be seen both calculating the force between parallel D3 branes
using the cylinder diagram of the type $0'$ string theory, or
calculating the Coleman-Weinberg potential in the field
 theory. Both calculations yield a no force result, and thus the scalar 
v.e.v. is indeed a modulus.
The existence of BPS states in field theory is due to the existence of such 
 objects in IIB string theory, or can be seen
by a field theory calculation
 in the planar limit. Finally, the fact that the large $N$ beta-function
 is zero is either due to S-duality of the IIB action or to S-duality
in \Nfour SYM.  

A more interesting question is whether some of these 
features survive at finite $N$. Since the theory is not supersymmetric and 
even the Bose-Fermi degeneracy is destroyed at finite $N$ (we have 
$8(N^2-1)$ bosons and $8(N^2-N)$ fermions), we do not expect a supersymmetric
behavior away from the planar limit. Clearly S-duality is broken, since
the $\beta$ function is no longer zero. However, some of these
features might be broken more softly than naively expected.

One example is the existence of a moduli space. The cylinder
 diagram yields a zero force between D3 branes, as a result
of the 
 balance between the exchange of the NS-NS fields $G_{\mu\nu}$ and $\Phi$
and of the R-R four form. The calculation of the Coleman-Weinberg one-loop 
effective potential yields the same result. The bosonic contribution
 to the potential is \cite{Zarembo,TZ} 
\beq 
V_{bos}(r) = {1\over 4\pi ^2} r^2 \log {r^2\over \Lambda ^2}\ ,
\eeq
 where $r$ is the v.e.v. of the
 scalar (the separation between the D-branes) and $\Lambda$ is some
 cutoff. The fermionic
 contribution is $V_{fer}(r) = - V_{bos}(r)$, and therefore 
the net force is zero, to
 this order. Note that, in contrast with the $SU(N)\times SU(N)$ gauge
 theory on the D3 branes of type 0B, where some directions were
 unstable and the gauge group was dynamically broken to its $U(1)$'s \cite{TZ},
 in the present case there is no ${1\over N}$ potential at tree level.
 In particular there is no instability
 and there seems to be a moduli space even at finite $N$. However,
 as we stated above,
 higher order perturbation theory will probably lift the vacuum energy
at generic points in the moduli space.

\section{Non-conformal large $N$ field theories}

The procedure described in the previous section is quite
general. Any supersymmetric theory with fermions in the adjoint or
bi-fundamental representations can be deformed into a
non-supersymmetric theory by changing the (color) representation of
the fermions,
in such a way that the large $N$ theory will converge to its parent 
supersymmetric theory.
The rules are: $adjoint \rightarrow \asymm + \basymm$, 
$(\fund,\bfund) \rightarrow (\bfund,\bfund)$ and
$(\bfund,\fund)\rightarrow (\fund,\fund)$. In particular,
non-conformal \Ntwo and \None theories can be used as parents.

Without the background of a string theory, we will not be able to say
much about the non-supersymmetric daughter theory. In fact, generally 
we expect a tachyonic instability at strong coupling\cite{klebanov}, and
for this reason we restrict ourself to theories that live
on D-branes with a non-tachyonic bulk theory.
Two such theories were constructed in section 2 (other cases were
recently discussed within the type I \cite{ADS,AADDS} and type IIB
\cite{uranga} frameworks).
 The orientifold projection of
 the type 0B theory on the $Z_2$ orbifold will converge in the planar limit to
the \Ntwo $SU(N)\times SU(M)$ gauge theory with hypermultiplets in the 
$(\fund,\bfund)$ and in the $(\bfund,\fund)$ representations. Thus, up
to ${1\over N}$ corrections, \Ntwo results apply to this model.

The second example is the $Z_3$ orbifold. It
 is interesting to take the limit when two of the gauge couplings of the
 theory are set to zero and the rank of these groups is equal. We also
take only one copy of the matter multiplet.
 In this limit, we will have an $SU(N_c)\times SU(N_f)\times SU(N_f)$
 theory with the non-supersymmetric field content of table 3 
\begin{table}[H]
\begin{displaymath}
\begin{array}{l c@{ }c@{ } c@{ } c}
& \multicolumn{1}{c@{\times}}{SU(N_c)}
& \multicolumn{1}{c@{\times}}{SU(N_f)}
& \multicolumn{1}{c@{}}{SU(N_f)} \\
\hline
\rm {vector}  & adj.  & 1 & 1\\
\rm {fermion} & \asymm + \basymm & 1 & 1 \\
\hline
 \rm {scalar} & \bfund & \fund & 1 \\
 \rm {scalar} & \fund & 1 & \bfund  \\
 \rm {fermion} & \fund & \fund & 1 \\
 \rm {fermion} & \bfund & 1 & \bfund \\
\end{array}
\end{displaymath}
\caption{The field content of an ``\None'' theory}
\label{electric}
\end{table}
In the large $N_c,N_f$ limit (with the ratio  ${N_f\over N_c}$ fixed), this
 theory converges to \None SQCD. Therefore, in the planar limit this theory
 is dual in the infra-red to a "magnetic" $SU(N_f-N_c)$ theory\cite{Seiberg}.
 It would be interesting if some sort of similar duality holds at finite $N$
as well. Note that, in contrast to the type 0 unprojected
 theory\cite{AK}, the one loop beta function receives a finite $N$
 correction (due to the orientifold)
\beq
b_1 = 3N_c - N_f - {4\over 3},
\eeq
and that the $U_R (1), U_R ^3 (1)$ anomalies match only in the planar limit. 
Therefore, the duality map should receive a ${1\over N}$ modifications.

\Acknowledgements

We are grateful to O. Aharony, C. Bachas, M. Bianchi, E. Gardi,
G. Grunberg, S. Kovacs, Y. Oz, A. Sagnotti and especially 
to I. Antoniadis and F. Hassan for stimulating discussions. 
C.A. acknowledges the Physics Departement of the University of Rome 
``Tor Vergata'' for the warm hospitality while this work was being
completed.
This research was supported in part by EEC under TMR contract 
ERBFMRX-CT96-0090.

\end{document}